\documentclass[10pt]{iopart}
\usepackage{iopams}
\usepackage{graphicx,epsfig}
\usepackage{dcolumn}
\usepackage{bm}

 \def\gtap{\mathrel{ \rlap{\raise 0.511ex \hbox{$>$}}{\lower 0.511ex
   \hbox{$\sim$}}}} 
\def\ltap{\mathrel{ \rlap{\raise 0.511ex
    \hbox{$<$}}{\lower 0.511ex \hbox{$\sim$}}}}

\newcommand{\apj}{ApJ}

\newcommand{\apjs}{ApJ Suppl.}

\newcommand{\mnras}{MNRAS}

\newcommand{\physrep}{Physics Reports}
\newcommand{\prd}{Phys. Rev. D}

\begin{document}
\title[Robust Neutrino Constraints from Low Redshift Observations]{Robust Neutrino Constraints by Combining Low Redshift Observations with the CMB}
\author{Beth A. Reid$^{1}$, Licia Verde $^{1,2,3}$, Raul Jimenez$^{1,2,3}$, Olga Mena$^4$}
$^{1}$ Institute for Sciences of the Cosmos (ICC), University of Barcelona, Barcelona 08028, Spain \\
$^{2}$ ICREA (Institucio Catalana de Recerca i Estudis Avancats) \\   
$^{3}$Theory group, Physics Department,  CERN, C-1211, Geneva 23, Switzerland\\ 
$^{4}$  Instituto de Fisica Corpuscular, IFIC, CSIC and Universidad de Valencia, Spain

\begin{abstract}
We illustrate how  recently improved low-redshift  cosmological measurements can tighten constraints on neutrino properties. In particular we  examine the impact of the assumed cosmological model on the constraints.  
We first consider the new HST $H_0 = 74.2 \pm 3.6$ measurement by Riess et al. (2009) and the $\sigma_8 (\Omega_m/0.25)^{0.41} = 0.832 \pm 0.033$ constraint from Rozo et al. (2009) derived from the SDSS maxBCG Cluster Catalog.  In a $\Lambda$CDM model and when combined with WMAP5 constraints, these low-redshift measurements constrain $\sum m_{\nu} < 0.4$ eV at the 95\% confidence level. 
This bound does not relax when allowing for the running of the spectral index or for primordial tensor perturbations.  When adding also Supernovae and BAO constraints, we obtain a 95\% upper limit of $\sum m_{\nu}<0.3$eV. We test the sensitivity of the neutrino mass constraint to the assumed expansion history by  both allowing a dark energy equation of state parameter $w\ne -1$ and by studying a model with coupling between dark energy and dark matter, which allows for variation in $w$, $\Omega_k$, and dark coupling strength $\xi$.  When combining CMB,  $H_0$ and the SDSS LRG halo power spectrum from  Reid et al. 2009, we find that in this very general model, $\sum m_{\nu} < 0.51$ eV with 95\% confidence.   If we allow the number of relativistic species $N_{\rm rel}$ to vary in a $\Lambda$CDM model with $\sum m_{\nu} = 0$, we find $N_{\rm rel} = 3.76^{+0.63}_{-0.68} (^{+1.38}_{-1.21})$ for the 68\% and 95\% confidence intervals.  We also report prior-independent constraints, which are in excellent agreement with the Bayesian constraints.
\end{abstract}
\section{INTRODUCTION}
Atmospheric and solar neutrino experiments have demonstrated that neutrinos have mass, implying a lower limit on the total mass of 0.056 eV \cite{lesgourgues/pastor:2006}.  Ongoing and future direct experiments will be sensitive to the absolute neutrino mass scale:  for instance, KATRIN will measure or constrain the electron neutrino mass to the 0.2 eV level.\footnote{{http://www-ik.fzk.de/tritium/motivation/sensitivity.html}}  A thermal neutrino relic component in the universe can impact both the expansion history and growth of structure.  Cosmological probes of neutrinos are therefore complementary to direct experiments and provide some of the tightest constraints on both the number of relativistic species present in the early universe and the sum of neutrino masses \cite{lesgourgues/pastor:2006}.  However, up to date, these constraints were dependent on the assumed cosmological model.  For example, bounds on the sum of the neutrino masses could be relaxed by, for example, allowing coupling in the dark sector  \cite{lavacca/bonometto/colombo:2009} or allowing hadronic axions \cite{hannestad/etal:2007}.  In this paper we  summarize how cosmic neutrinos affect cosmic history, examine how recently improved  cosmological measurements of  the low redshift universe can help constrain these neutrino properties, and scrutinize the impact of our assumptions about the cosmological model.

In the standard model for particle physics  there are three massless neutrino species with weak interactions.
Neutrinos decouple early in cosmic history, when the temperature was $T \sim 1$ MeV, and thereafter contribute to the relativistic energy density with an effective number of species, $N_{\rm eff} = 3.046$ \cite{lesgourgues/pastor:2006}. Cosmology is sensitive to the physical energy density in relativistic particles in the early universe $\omega_{\rm rel}$ which, in the standard model (for cosmology), includes only photons and neutrinos $\omega_{\rm rel}=\omega_{\gamma}+N_{\rm eff}\omega_{\nu}$ where $\omega_{\gamma}$ is the energy density in photons and $\omega_{\nu}$ is 
the energy density in one active neutrino. As $\omega_{\gamma}$ is extremely well constrained, $\omega_{\rm rel}$ can be used to constrain neutrino properties;  deviations from  $N_{\rm eff} = 3.046$  would signal non-standard neutrino features or additional relativistic relics.  $N_{\rm eff}$ impacts the big bang nucleosynthesis epoch through its effect on the expansion rate.  Therefore measurements of the primordial abundance of light elements can constrain $N_{\rm eff}$ \cite{cyburt:2004, cyburt/etal:2005, steigman:2007, iocco/etal:2009}.  These constraints rely on physics at the time of big-bang nucleosynthesis ($T\sim $ MeV), and in several non-standard models the energy density in relativistic species can change  at later time (e.g. at the last scattering, $T \sim$ eV). 
Free-streaming relativistic particles affect the Cosmic Microwave Background (CMB) through their relativistic energy density, which alters the epoch of matter-radiation equality, and through their anisotropic stress \cite{trotta/melchiorri:2005,komatsu/etal:2009}.  Because the redshift of matter-radiation equality is well constrained by the ratio of the third to first CMB peak height, the first effect defines a degeneracy between $\Omega_m h^2$ and $N_{\rm eff}$ \cite{bowen/etal:2002}:
\begin{equation}
\label{zequal}
1+z_{eq} = \frac{\Omega_m h^2}{\Omega_{\gamma} h^2}\frac{1}{1+0.2271 N_{\rm eff}}.
\end{equation}
Because of the second effect, WMAP5 data favors $N_{\rm eff} = 3.046$ over $N_{\rm eff} = 0$ at the 99.5\% confidence level \cite{dunkley/etal:2009}.

Neutrinos with mass $\ltap1$ eV become non-relativistic after the epoch of recombination probed by the CMB, so that allowing massive neutrinos alters matter-radiation equality for fixed $\Omega_m h^2$. 
Their radiation-like behavior at early times  changes the expansion rate, shifting the peak positions, but this is somewhat degenerate with other cosmological parameters. Therefore, WMAP5 alone constrains $\sum m_{\nu} < 1.3$ eV at the 95\% confidence in a flat $\Lambda$CDM universe, and  this constraint relaxes to $\sum m_{\nu} < 1.5$ eV for a flat $w$CDM universe, assuming $N_{eff} = 3.046$ \cite{dunkley/etal:2009}.  Figure 17 of \cite{komatsu/etal:2009} shows that there remain degeneracies between $\sum m_{\nu}$ and both $H_0$ and $\sigma_8$; in this paper we show that currently available low-redshift measurements can break these degeneracies, yielding improved  and robust constraints on $\sum m_{\nu}$.

After the neutrinos become non-relativistic, their free-streaming damps power on small scales and therefore modifies the matter power spectrum in the low redshift universe. The latest large-scale structure constraints from the Sloan Digital Sky Survey Luminous Red Galaxy sample in combination with WMAP5 yield $\sum m_{\nu} < 0.62$ eV in a flat $\Lambda$CDM model with $N_{\rm eff} = 3.046$ \cite{reid/etal:2009}.  In addition, because the measurement of the matter power spectrum provides a separate constraint on the transfer function, this data combination also yields a constraint on $N_{\rm eff} = 4.8^{+1.8}_{-1.7}$ in a flat $\Lambda$CDM cosmology with no massive neutrinos.  We will show that the combination of low redshift data used in this paper provides tighter neutrino constraints in both of these cases.

Relaxing the assumptions about the cosmic expansion history, e.g., requiring $w=-1$ vs. allowing $w$ as a free parameter \cite{komatsu/etal:2009,Verde/etal:2003} or allowing coupling in the dark sector can relax constraints on neutrino masses \cite{lavacca/bonometto/colombo:2009}.  As an example we study constraints on neutrino masses in the model of \cite{gavela/etal:2009}, allowing $\Omega_k$, $w$, and coupling strength $\xi$ to vary simultaneously.  

In Section~\ref{whichconstraints} we discuss the low-redshift observational constraints we will make use of to improve neutrino constraints, paying careful attention to how the constraints generalize to the models we consider.  In Sections \ref{sumneutrinomass} and \ref{nrelsec} we present new constraints on neutrino masses and the effective number of relativistic species at matter-radiation equality.  We also make a detailed comparison with results from different cosmological probes available in the literature.  In Section~\ref{conc} we summarize our findings and argue that they are robust to a wide variety of alterations of the cosmological model.

\section{DATA AND METHODS}
\label{whichconstraints}
In subsections \ref{genH0} and \ref{genmaxbcg} we discuss the low redshift constraints we make use of in our analysis, paying particular attention to their generalization to models not initially considered in their analyses.  In subsection \ref{method} we review our standard Bayesian methodology for computing parameter constraints and upper limits, and introduce the profile likelihood as a means to check the dependence on these constraints on our priors.
\subsection{$H_0$}
\label{genH0}
Ref.~\cite{riess/etal:2009} presents a redetermination of the Hubble constant that makes use of Cepheid variables in SN Ia host galaxies and in the ``maser galaxy'' NGC 4258, along with an expanded SN Ia sample at $z < 0.1$.  We reinterpret the reported constraint, $H_0 = 74.2 \pm 3.6$ km s$^{-1}$ Mpc$^{-1}$, as a constraint on the inverse luminosity distance at the effective redshift $z=0.04$ (Riess, private communication).  This correctly accounts for the slight dependence on $w$ shown in Figure 14 of \cite{riess/etal:2009}, and generalizes the constraint to an arbitrary cosmology.  In practice, when combining with other data sets, the cosmology dependence of this constraint is quite small.
\subsection{maxBCG}
\label{genmaxbcg}
Ref.~\cite{rozo/etal:2009} derives cosmological constraints from the maxBCG cluster catalog \cite{koester/etal:2007}, which identifies spatial overdensities of bright red galaxies in SDSS DR4+ imaging data and results in a nearly volume-limited catalog with photometric redshifts in the range $z=0.1$ to $z=0.3$.  MaxBCG clusters are classified by $N_{200}$, the number of red-sequence galaxies within a scaled radius such that the average galaxy overdensity interior to that radius is 200 times the mean galaxy density.  The group multiplicity as a function of $N_{200}$ is one of two primary observable inputs to the maxBCG posterior.  The second is the mean mass measured via gravitational lensing for 5 bins in $N_{200}$ from \cite{johnston/etal:2007}.

The analysis of Ref.~\cite{rozo/etal:2009} assumes a flat $\Lambda$CDM cosmology with $N_{eff} = 3.046$ massless neutrinos, where $\sigma_8$ and $\Omega_m$ are allowed to vary, and $h=0.7$, $n_s = 0.96$, and $\Omega_b h^2 = 0.02273$ are held fixed.  There are four nuisance parameters: two to define the mean power law relationship between $N_{200}$ and halo mass $M$, one to quantify the size of the log-normal scatter about this relation, and one for the lensing mass bias parameter.  The principal cosmological constraint derived from the maxBCG catalog is
\begin{equation}
\label{maxbcgconstraint}
\sigma_8 (\Omega_m/0.25)^{0.41} = 0.832 \pm 0.033.
\end{equation}
In Section 5.2.1, Ref.~\cite{rozo/etal:2009} addresses the limitations of their assumptions on cosmological parameters.  They find that using a Gaussian prior on $h = 0.7 \pm 0.1$ and $n_s = 0.96 \pm 0.05$ does not alter their $\sigma_8\Omega_m^{0.41}$ constraint.  In addition, for a prior with $\sum m_{\nu} < 1$ eV, the $\sigma_8\Omega_m^{0.41}$ constraint is not altered.  Massive neutrinos suppress the linear power spectrum, but because $\sigma_8$ corresponds to the mass scale to which maxBCG is most sensitive, the impact of the alteration in power spectrum shape is minimal on the final constraints.  In principle, massive neutrinos also alter growth in the nonlinear regime.  Ref.~\cite{brandbyge/etal:2008} shows that power suppression is enhanced in the nonlinear regime for massive neutrinos, implying that the upper bound reported in the present paper is robust.  Therefore, since the range of total neutrino mass we explore in this paper is well below $1$ eV, we can safely apply the maxBCG constraint.  In the case of massive neutrinos, we also study the maxBCG constraint in combination with WMAP5 for a $\Lambda$CDM model including spectral index running.  When we evaluate the spectral index $n_{s}(k)$ for all $k \leq 0.4 \; h$/Mpc from the WMAP5 constraints in this model, the value falls between $0.91$ and $1$ for 68\% of the time.  This range satisfies the stated maxBCG tolerance to variation in the spectral index, and therefore we may apply the maxBCG constraint unaltered in the models we study including massive neutrinos.

The maxBCG analysis did not directly address models in which $N_{\rm eff}$ is varied.  However, we have verified that within the space of parameters allowed by WMAP5 and the $H_0$ constraint of \cite{riess/etal:2009} in a $\Lambda$CDM+$N_{\rm eff}$ model, the impact of $N_{\rm eff}$ on the shape of the linear power spectrum can be compensated by a change in $n_s$ that is within the maxBCG stated tolerances.  Therefore we may apply the constraint on the power spectrum amplitude through $\sigma_8$ in Equation \ref{maxbcgconstraint} without modification when $N_{\rm eff}$ is varied.
\subsection{Markov Chain Monte Carlo and Profile Likelihoods}
\label{method}
We use the standard Markov Chain Monte Carlo (MCMC) scheme to probe the posterior distribution for several different cosmological models \cite{dunkley/etal:2009,spergel/etal:2003,spergel/etal:2007}.  The publicly available WMAP5 chains which we resample have uniform priors on the parameters $\{\Omega_b h^2, \Omega_c h^2, \Omega_\Lambda, \tau, n_s, \Delta_{\cal R}^2, A_{SZ}\}$, the baryon density, cold dark matter density, dark energy density, reionization optical depth, scalar spectral index at $k_0 = 0.002$ Mpc$^{-1}$, amplitude of curvature perturbations at $k_0$, and SZ marginalization factor.  The MCMC chain we produce for the the dark coupling model of \cite{gavela/etal:2009} uses the same parameters, but replaces $\Omega_\Lambda$ with the parameter $\theta$, the ratio of the approximate sound horizon to the angular diameter distance, which is used by CosmoMC \cite{lewis/bridle:2002}.

For a more ``frequentist'' statistic, the profile likelihood can be computed from the MCMC chain \cite{hamann/etal:2007}. Since we have 7 or more uninteresting parameters and only one parameter for which we want to report constraints (say, $\beta$), for each value of $\beta$ we find the maximum likelihood value $L_{\beta}$, regardless of the values assumed by other parameters. We then consider $\ln L_{\beta}/L_{max}$ as a function of the model parameter $\beta$, or function thereof; here $L_{max}$ is the maximum likelihood over the entire chain.  We examine the profile likelihood for both $\sum m_{\nu}$ (Section \ref{profilelike}) and $N_{\rm eff}$ (Section \ref{neffprofilelike}).  As long as the posterior distribution is singly peaked, this profile should be independent of the Bayesian priors; but note that the profile likelihood will become noisier as $\ln L/L_{max}$ decreases, since the MCMC method means the chains do not densely sample low $\ln L/L_{max}$ regions.
\section{CONSTRAINTS ON THE SUM OF NEUTRINO MASSES}
\label{sumneutrinomass}
We begin in Section \ref{mnusec} by studying the $\Lambda$CDM model and a few one-parameter extensions explored in the WMAP5 analysis.  In section \ref{darkcouplingresults} we study the dark coupling model of \cite{gavela/etal:2009} as an example of a model which significantly relaxes the model assumptions about the expansion history by varying three additional parameters.  We consider the profile likelihood to study the dependence of our constraints on our cosmological priors and check the consistency of the data sets we examine.  Finally we compare with previous constraints on $\sum m_{\nu}$ in the literature using a variety of data sets.
\subsection{$\Lambda$CDM and one-parameter extensions}
\label{mnusec}
\begin{figure}
  \centering
 \includegraphics[width=0.7\columnwidth,height=0.6\columnwidth]{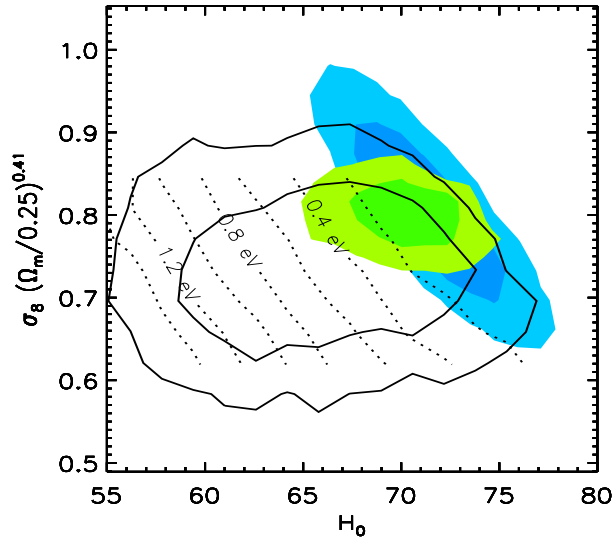}
  \caption{\label{fig:mnulcdm}  $\Lambda$CDM WMAP5-only constraints with $\sum m_{\nu} = 0$ are shown in blue.  The black transparent contours show how the constraints on $\sigma_8 (\Omega_m/0.25)^{0.41}$ and $H_0$ degrade when $\sum m_{\nu}$ is left as  a free parameter.  The dotted lines show contours on which $\left < \sum m_{\nu} \right > = 0.2, 0.4, 0.6, ..., 1.4$ eV for the WMAP5-only posterior distribution.  The green shows the constraints when maxBCG and $H_0$ constraints are also included.  The 95\% confidence upper limit is reduced from 1.3 eV to 0.4 eV.}
\end{figure}
\begin{figure}
  \centering
\includegraphics[width=0.7\columnwidth,height=0.6\columnwidth]{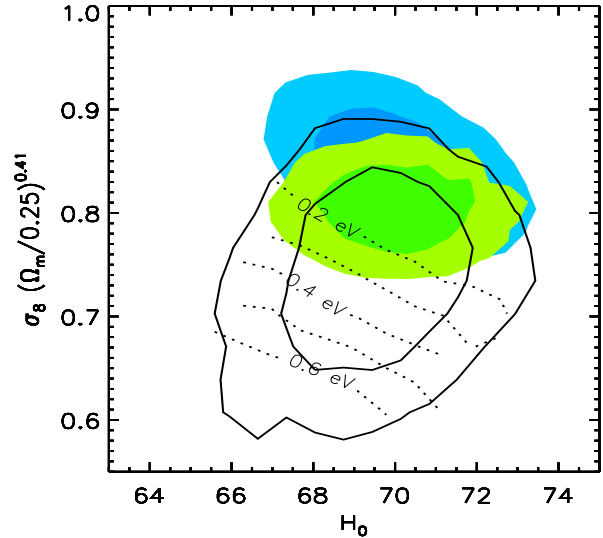}
  \caption{\label{fig:mnulcdmbaosn} Same as Figure \ref{fig:mnulcdm} but for the $\Lambda$CDM model with base dataset WMAP$+$BAO$+$SN.  The dotted lines show contours on which $\left < \sum m_{\nu} \right > = 0.2, 0.3, 0.4, 0.5, 0.6$ eV for the WMAP$+$BAO$+$SN posterior distribution.}
\end{figure}
\begin{table*}
\begin{center}
\begin{tabular}{llcccc}
\multicolumn{6}{c}{Bayesian and Frequentist upper 95\% C.L. bound on $\sum m_{\nu}$}\\
\hline
model & base dataset & -- & $+$maxBCG & $+H_0$ & $+$maxBCG$+H_0$\\
\hline
$\Lambda$CDM & WMAP5 &  1.3 & 1.1 & 0.59 & 0.40 \\
$\Lambda$CDM & WMAP5$+$BAO$+$SN &  0.67 & 0.35 & 0.59 & 0.31 \\
$\Lambda$CDM $+\alpha $ & WMAP5 & 1.34 & 1.25 & 0.54 & 0.39 \\
$\Lambda$CDM $+r$ & WMAP5 & 1.36 & 1.18 & 0.83 & 0.40 \\
$w$CDM & WMAP5$+$BAO$+$SN & 0.80 & 0.52 & 0.72 & 0.47 \\
 dark coupling & WMAP5$+\hat{P}_{halo}(k)+$SN & - & - & 0.51 & - \\
 \hline
$\Lambda$CDM & WMAP5 &  1.3 & 0.82 & 0.50 & 0.33 \\
$\Lambda$CDM & WMAP5$+$BAO$+$SN &  0.66 & 0.32 & 0.56 & 0.30 \\
$\Lambda$CDM $+\alpha $ & WMAP5 &  1.28 & 1.15 & 0.63 & 0.43 \\
$\Lambda$CDM $+r$ & WMAP5 &  1.23 & 0.86 & 0.72 & 0.30 \\
$w$CDM & WMAP5$+$BAO$+$SN & 0.80 & 0.44 & 0.74 & 0.44 \\
 dark coupling & WMAP5$+\hat{P}_{halo}(k)+$SN & - & - & 0.55 & - \\
\end{tabular}
\caption{\label{table:mnu} 95\% upper limit on the sum of the neutrino masses $\sum m_{\nu}$ in eV.  In the upper half of the table the limits are Bayesian, i.e. they represent the $\sum m_{\nu}$ value below which 95\% of the weighted points in the MCMC chain lay.  The lower half of the table approximates  a frequentist interpretation: we report the largest value of $\sum m_{\nu}$ in the chain for which $\ln L_{\sum m_{\nu}}/L_{max} \ge - 2$.   For the first five models we use the WMAP5 public MCMC chains to compute the upper bounds; these chains use WMAP5 year data, and where noted, BAO constraints from  \cite{percival/etal:2007} and the Union supernova sample \cite{kowalski/etal:2008}.  Column 1 specifies the cosmological model assumed.  All five of these models assume flatness, and $w$ is assumed constant if varied.  $\alpha \equiv dn_s/d ln k$ indicates running in the primordial power spectrum and $r$ indicates allowing for primordial tensor modes.  Column 2 specifies the original datasets used to determine the constraints shown in Column 3, and Columns 4-6 give the resulting upper bounds after applying additional observational constraints from maxBCG (Column 4), $H_0$, (Column 5), and both of these (Column 6).  In the last column we include the constraint from Section ~\ref{darkcouplingresults} on the model of Ref.~\cite{gavela/etal:2009}.  In that model, curvature, $w$, and dark coupling $\xi$ are simultaneously varied.}
\end{center}
\end{table*}
Figure \ref{fig:mnulcdm} shows the WMAP5-only constraints in a $\Lambda$CDM model when $\sum m_{\nu} = 0$ (blue contours).  The solid black contours show that when $\sum m_{\nu}$ is allowed to vary, the constraints on $H_0$ and $\sigma_8 (\Omega_m/0.25)^{0.41}$ weaken substantially.  While both the maxBCG and $H_0$ constraints are in good agreement with the $\Lambda$CDM values, they effectively exclude high neutrino masses (green contours).  Adding both constraints reduces the 95\% confidence limit on $\sum m_{\nu}$ from 1.34 to 0.4 in a flat $\Lambda$CDM cosmology.  Figure \ref{fig:mnulcdmbaosn} shows the same scenario, but when BAO and SN data have also been combined with WMAP5.  Since those probes break the degeneracy between $H_0$ and $\Omega_m$ in $\Lambda$CDM cosmology, $H_0$ is well constrained, almost independently of $\sum m_{\nu}$.  In this case the maxBCG result drives the reduction of the 95\% confidence bound from 0.67 eV to 0.31 eV.

In Table~\ref{table:mnu} we present these and several other one dimensional constraints on $\sum m_{\nu}$ for the $\Lambda$CDM model, as well as several other data and model variants. 
In the upper half of the table the limits are Bayesian, i.e. they represent the $\sum m_{\nu}$ value below which 95\% of the weighted points in the MCMC chain lay.  The lower half of the table approximates  a frequentist interpretation (a more detailed discussion of this is in sec.~\ref{profilelike}): we report the largest value of $\sum m_{\nu}$ in the chain for which $\ln L_{\sum m_{\nu}}/L_{max} \ge - 2$. While this limit can be interpreted in terms of numbers of standard deviations only for Gaussian distributions, it is prior-independent. We have checked that    the $\sum m_{\nu}$ distribution is well approximated by a  Gaussian.

 The constraint on $\sum m_{\nu}$ from WMAP5$+$maxBCG$+H_0$ does not relax when running or tensors are allowed.  The constraint relaxes slightly to $\sum m_{\nu} < 0.47$ eV at the 95\% confidence level when $w$ is allowed to vary, but BAO and SN constraints are also included.  The constraint relaxes because the maxBCG constraint - $\sum m_{\nu}$ degeneracy is less steep for this model, since changing $w$ also impacts the growth of structure.  These examples indicate that when the expansion history is fixed, $\sum m_{\nu} < 0.4$ eV (95\% confidence) is a robust upper limit; changes in the primordial fluctuations that are consistent with WMAP5 do not impact this bound.  However, when the expansion history  is altered at late times ($w \ne -1$ but constant), this bound relaxes slightly.  As an extreme example of this, we study the dark coupling model of \cite{gavela/etal:2009} next.
\subsection{Dark coupling and massive neutrinos}
\label{darkcouplingresults}
In this subsection we consider the particular model of coupling in the dark sector proposed by Ref.~\cite{gavela/etal:2009}. In their model the densities of dark matter and of dark energy do not evolve independently as the universe expands but are coupled, with an interaction strength parameterized by $\xi$ as follows:
\begin{eqnarray}
\nabla_{\mu} T^{\mu}_{(dm)\nu} & = & \xi H\rho_{de} u_{\nu}^{dm} \label{coupling1}\\
\nabla_{\mu} T^{\mu}_{(de)\nu} & = & -\xi H\rho_{de} u_{\nu}^{dm} \label{coupling2},
\end{eqnarray}
where $T^{\mu}_{(dm)\nu}$ and $T^{\mu}_{(de)\nu} $ are the energy momentum tensors for the dark matter and dark energy components, respectively, $u^{dm}$ denotes the dark matter four velocity and $H$ the Hubble parameter.
The coupling strength  $\xi$ alters the expansion history through changes in the matter and dark energy density evolution \cite{gavela/etal:2009}:
\begin{eqnarray}
\rho_{dm}(a) & = & \rho_{dm}^{0} a^{-3} + \rho_{de}^{0} \frac{\xi}{3w_{de}^{\rm eff}} (1 - a^{w_{de}^{\rm eff}})a^{-3} \label{coupling3}\\
\rho_{de}(a) & = & \rho_{de}^{0} a^{-3(1+w_{de}^{\rm eff})} \label{coupling4}
\end{eqnarray}
where $w_{de}^{\rm eff} = w+\xi/3$, the superscript $0$ denotes the present-day quantity, and $a$ denotes the scale factor.
We adopt this model as a suitable example for which to study whether the constraint on $\sum m_{\nu}$ relaxes when several free parameters are allowed to simultaneously vary.  It was shown by Ref.~\cite{Kristiansen/LaVacca/Colombo/Mainini/Bonometto:2009, LaVaccaKristiansen/Colombo/Mainini/Bonometto:2009,LaVacca/Kristiansen:2009} that allowing for  a coupling in the dark sector dramatically relaxes  cosmological constraints on neutrino mass. Ref.~\cite{gavela/etal:2009}  presents a slightly different model for dark coupling, which improves on existing models in the literature as it  does not suffer from instabilities; they present  cosmological constraints on this dark coupling parameterization in a very general model, allowing the curvature, equation of state of dark energy $w \geq -1$, total neutrino mass, and the amplitude of dark coupling $\xi < 0$ between dark matter and dark energy to vary along with the standard $\Lambda$CDM cosmological parameters.  These additional parameters allow the expansion history and growth of structure to deviate appreciably from the $\Lambda$CDM assumption. We will explore the cosmological constraints on the coupling  strength $\xi$ in a future work alongside this model's non-standard growth of structure; because of the large deviations from the  $\Lambda$CDM    growth of structure, the maxBCG cluster constraint, which is based on the standard spherical collapse-based correspondence between the linear matter power spectrum and the resulting halo mass function in $\Lambda$CDM cosmologies, cannot be applied to the \cite{gavela/etal:2009} model.  
For this model, we therefore substitute the maxBCG constraints  with  the SDSS DR7 Luminous Red Galaxies $\hat{P}_{halo}(k)$ constraint of \cite{reid/etal:2009}  as a robust low redshift probe of $\sum m_{\nu}$.

\begin{figure}
  \centering  
 \includegraphics[width=0.75\columnwidth,height=0.65\columnwidth]{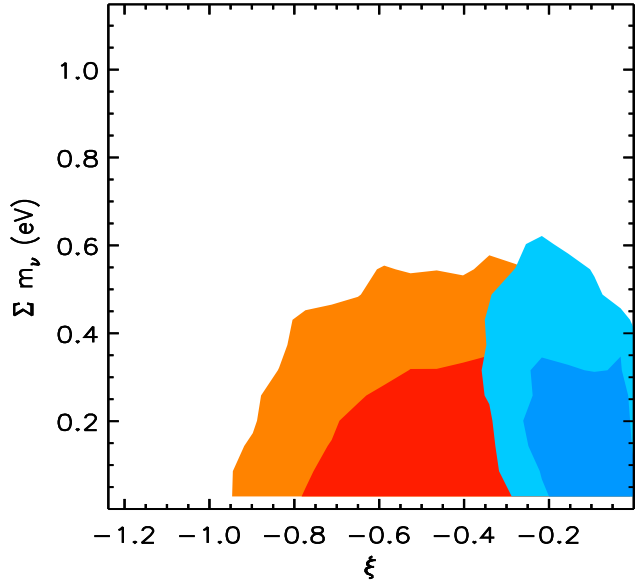}
  \caption{\label{fig:mcmc} Two-dimensional constraints in $\xi-\sum m_{\nu}$ in the dark coupling model of ref. \cite{gavela/etal:2009}.  The red contours are the results from the CMB+ $H_0$ \cite{riess/etal:2009}+ $\hat{P}_{halo}(k)$ \cite{reid/etal:2009}+ Union SN \cite{kowalski/etal:2008}.  With this data-set combination there is no longer a degeneracy between coupling and neutrino mass. As discussed in Section ~\ref{darkcouplingresults}, the strong changes in the growth of structure caused by the dark coupling prevent us from applying the maxBCG constraint to this model.  However, the blue contours, which also include the maxBCG constraint, demonstrate that the constraint on $\sum m_{\nu}$ is not tightened even if this additional constraint could be included.}
\end{figure}

We study the dark coupling model using the MCMC method as implemented in the COSMOMC package \cite{lewis/bridle:2002}.  Several datasets must be included to avoid long degeneracies in the parameter space.  We include WMAP5 data along with several low redshift datasets:  the new $H_0$ constraint \cite{riess/etal:2009},  SDSS DR7 Luminous Red Galaxies halo power spectrum $\hat{P}_{halo}(k)$ \cite{reid/etal:2009}, and  Union SN sample \cite{kowalski/etal:2008}.  The only important prior we enforce is $\Omega_{dm} h^2 > 0.001$; surprisingly, even models consistent with these four cosmological probes can have negligible $\Omega_{dm} h^2$ today since the coupling accelerates the transition to dark energy domination.  Note that massive neutrinos suppress $\hat{P}_{halo}(k)$ in a scale-dependent fashion, thus, by including the $\hat{P}_{halo}(k)$ measurements, one expects to be able to disentangle the effects of neutrino mass from the effects of  alterations of the late-time expansion history.  Figure \ref{fig:mcmc} shows constraints from these four observational probes (red contours):  $\sum m_{\nu}$ is not degenerate with dark coupling $\xi$, and the resulting one dimensional constraint on $\sum m_{\nu}$ is 0.51 eV.  For completeness, we also show in Figure ~\ref{fig:mcmc} how the constraints would tighten {\em if} the maxBCG constraint were applicable (blue contours).  The constraint on the neutrino mass is unaffected.

\subsection{Comparison with constraints from the profile likelihood}
\label{profilelike}
\begin{figure}
  \centering
 \resizebox{0.7\columnwidth}{!}{\includegraphics{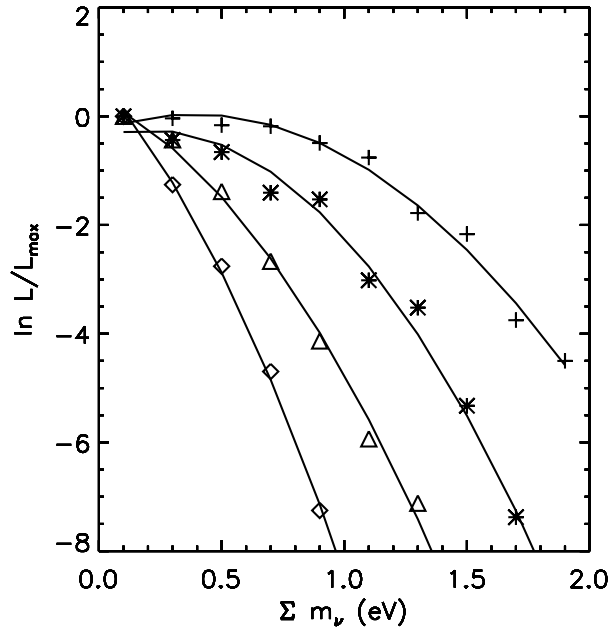}}
  \caption{\label{fig:profilelike}  The profile likelihood defined in Section \ref{method} in bins of $\Delta (\sum m_{\nu}) = 0.2$ eV for the $\Lambda$CDM model with WMAP5 only (crosses), WMAP5+maxBCG (stars), WMAP5+$H_0$ (triangles), and WMAP5+maxBCG+$H_0$ (diamonds).  The black curves overlay a quadratic fit to these points, illustrating that a Gaussian curve provides a good fit to this one-dimensional distribution.}
\end{figure}

We first verify that all three primary datasets are consistent with each other in the $\Lambda$CDM cosmology by comparing the maximum likelihood value of the WMAP5 data alone with the maximum likelihood after adding the additional observational constraints.  We find that $\ln L_{max}$ decreases by 0.5 with the addition of the two independent Gaussian priors from the $H_0$ and maxBCG constraints; the decrease in $\ln L_{max}$ is the same when we add massive neutrinos to the model.  Therefore, our tight constraints are not arising from tension between the datasets we are using.  To compute the profile likelihood as described in Section \ref{method}, we find the maximum likelihood value over the MCMC chain in bins of $\sum m_{\nu}$.  Figure \ref{fig:profilelike} shows that this function is well-fit by a Gaussian in each case.  We can therefore  tentatively interpret the largest $\sum m_{\nu}$ value for which $\ln L/L_{max} \geq -2$ as a prior-independent  $\sim 2\sigma$ upper bound on $\sum m_{\nu}$.  We report this value in the bottom half of Table \ref{table:mnu}; in general we find excellent agreement with the Bayesian results.  
\subsection{Comparison with other $\sum m_{\nu}$ constraints}
\label{comparemnuconstraints}
In Table \ref{table:mnuother} we compile other $\sum m_{\nu}$ constraints available in the literature and compare with the results we have found.  WMAP5 provides a robust upper limit in a $\Lambda$CDM model, $\sum m_{\nu} < 1.3$ eV; allowing $w \neq -1$ relaxes the constraint only slightly to $\sum m_{\nu} < 1.5$ eV.  In a study of WMAP3 data, Ref.~\cite{Kristiansen/Eriksen/Elgaroy:2006} find that correcting the likelihood approximation on large angular scales as well as accounting for an over-subtraction of point sources affecting small angular scales, the WMAP3-only 95\% upper bound improved from 1.9 to 1.57 eV.  $\sum m_{\nu}$ remains degenerate with both $H_0$ and $\sigma_8$ within the WMAP5 constraint.  Using BAO and SN reduces the degeneracy with $H_0$, reducing the limit to 0.67 eV ($w=-1$) or 0.8 eV when $w$ is also allowed to vary.  Using  both the $k$-dependent suppression of power and BAO feature in the matter power spectrum derived from the SDSS Luminous Red Galaxy sample, Ref.~\cite{reid/etal:2009} finds $\sum m_{\nu} < 0.62$ eV when combined with WMAP5 for $w=-1$.  That analysis improves the nonlinear modeling by removing the additional shot noise contribution from galaxies occupying the same dark matter halo and marginalizing over the remaining modeling uncertainties.  Ref.~\cite{hamann/etal:2008} compares the impact of two nonlinear modeling approaches on the cosmological inferences for several extensions of standard $\Lambda$CDM cosmology, and advocates marginalizing over an additional shot noise term.

Additional progress requires probing the amplitude of mass fluctuations.  This has been done with several samples probing the cluster mass function.  Ref.~\cite{vikhlinin/etal:2009} derives a low and high redshift cluster mass function from {\em Chandra} observations of clusters in two ROSAT surveys.  Their measurement provides a constraint from the normalization of the local mass function $\sigma_8 (\Omega_m/0.25)^{0.47} = 0.813 \pm 0.013 \pm 0.024$ (statistical and systematic errors, respectively); this is similar to and in agreement with the maxBCG constraint we have used in our analysis.  Additional cosmological information is available by comparing the high and low redshift mass functions.  Their tighter statistical constraint on the local mass function normalization translates to tighter constraints on $\sum m_{\nu} < 0.33$ eV, but with the caveat that $\sim 0.1$ eV of systematic error has not been included.  Moreover, Ref.~\cite{mantz/etal:2009} points out that the analysis of \cite{vikhlinin/etal:2009} does not fit for both the cluster scaling relations and cosmological parameters self-consistently.

Ref.~\cite{kristiansen/elgaroy/dahle:2007} uses the cluster mass function from weak lensing to derive a similar cluster normalization constraint: $\sigma_8 (\Omega_m/0.25)^{0.37} = 0.72^{+0.04}_{-0.05}$.  Because of both the larger errors, lower central value, and combination with WMAP3 rather than WMAP5, they find a relatively weak constraint, $\sum m_{\nu} < 1.43$ eV.

Because so far only upper limits can be obtained from cosmology on the neutrino masses, the tightness of the $\sum m_{\nu}$ constraints depends drastically on where the lower-redshift $\sigma_8$ determination intersects the CMB  $\sum m_{\nu}-\sigma_8$ degeneracy.
An example of this sensitivity is the constraint obtained  including the Ly-$\alpha$ forest power spectrum (LYA) measurements \cite{seljak/slosar/mcdonald:2006}.  In that work,  $\sim 2\sigma$ tension between the power spectrum amplitudes preferred by WMAP3 and LYA when $\sum m_{\nu} = 0$ was reported, with LYA preferring a larger amplitude.  This  translates into an extremely tight constraint on $\sum m_{\nu} < 0.17$ eV when LYA is combined with many other observational probes (WMAP3, SN, SDSS BAO, SDSS and 2dF $P(k)$, and small scale CMB experiments).
With updated datasets for WMAP and SN but including fewer datasets altogether, and in addition allowing for running of the spectral index, we find that LYA data still reduces the constraint to $\sum m_{\nu} < 0.28$ eV.  Moreover, Ref.~\cite{bolton/etal:2008} finds possible evidence for an inverted temperature-density relation, which would  change the interpretation of the LYA power spectrum measurement and significantly weaken the Ref.~\cite{seljak/slosar/mcdonald:2006} bound.

Table~\ref{table:mnu} demonstrated the improvement on $\sum m_{\nu}$ when tight Hubble constant measurements are included \cite{riess/etal:2009}. Refs.~\cite{stern/etal:2009,simon/etal:2005} use the stars in passively-evolving red galaxies as cosmic chronometers to obtain a cosmological-model independent measurement of  $H(z)$ from $z=0$ to $z\sim 1$; adding these additional constraints on $H(z)$ also reduces the constraint on $\sum m_{\nu}$ to 0.5 eV.  Weak lensing provides additional constraints on the amplitude of fluctuations on small scales which can be used to constrain $\sum m_{\nu}$; both Ref.~\cite{ichiki/takada/takahashi:2009} and Ref.~\cite{tereno/etal:2009} find a limit of $\sim 0.54$ eV.  Finally, there have also been constraints derived from luminosity-dependent biasing \cite{seljak/etal:2005, debernardis/etal:2008}.  The most recent work by \cite{debernardis/etal:2008} combines WMAP5 with the \cite{tegmark/etal:2006} LRG $P(k)$ and luminosity-dependent bias measurements to find $\sum m_{\nu} < 0.28$ eV in $\Lambda$CDM; this constraint relaxes to $0.59$ eV in a $w$CDM cosmology.  While the modeling procedure can be quite complex, this approach is certainly competitive with other probes.

In conclusion, WMAP5 provides a robust 95\% confidence upper bound of $\sim 1.5$ eV.  Probes of the redshift-distance or $H(z)$ relation break key degeneracies and reduce the constraints to $\sim 0.5-0.8$ eV.  Similarly, adding the  measurement of the galaxy power spectrum shape for SDSS  DR7 \cite{reid/etal:2009} reduces the bound to $0.62$ eV.  Beyond this, probes of the amplitude of clustering $\sigma_8$ through measurements of the cluster mass function, weak lensing, LYA, or luminosity-dependent clustering are necessary.  The most optimistic interpretation of the currently available data reduces the bound to $\sim 0.3$ eV.  All of these methods require more complicated modeling and need further quantification of the remaining systematics.

\begin{figure}
  \centering
 \resizebox{0.9\columnwidth}{!}{\includegraphics{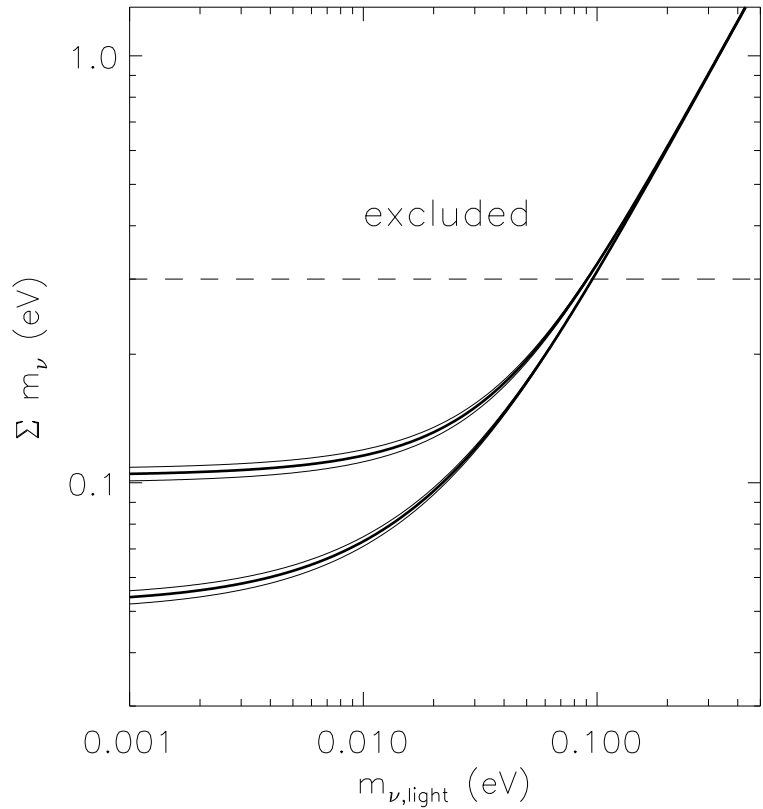}}
  \caption{\label{fig:hierarchy}  $\sum m_{\nu}$ vs. the mass of the lightest neutrino in the normal and inverted hierarchies.}
\end{figure}

From current solar and atmospheric constraints on neutrino mass differences $\Delta m^2$, three types of neutrino mass spectra are allowed:
the normal hierarchy ($m_1\ll m_2 \ll m_3$), the inverted hierarchy ($m_3\ll m_1 \simeq m_2$) and the quasi-degenerate hierarchy ($m_1\sim m_2 \sim m_3$). Constraints on the absolute neutrino mass scale would greatly help in distinguishing there three cases as
indicated in Figure \ref{fig:hierarchy}. Direct neutrino mass searches based on tritium beta decay so far have imposed an upper limit of $2.3$ eV at the $95\%$ CL~\cite{Kraus:2004zw}, but to  distinguish e.g. the quasi-degenerate hierarchy from the other cases a mass sensitivity  of $\sim 0.2$ eV is needed.
The neutrino mass hierarchy is one of the unknowns within the neutrino sector and its extraction would be crucial for determining the neutrino character (Dirac or Majorana) and measuring leptonic CP violation. A determination of the neutrino mass hierarchy would also rule out some neutrino mass models based on lepton flavor symmetries and/or grand unification schemes~\cite{Albright:2006cw}.
If future cosmological data finds no evidence for a neutrino mass down to $0.1$ eV, that would imply that the neutrino mass hierarchy is normal, allowing, in principle, a clean measurement of CP violation in the leptonic sector (provided the mixing angle $\theta_{13}$ is not very small), see Ref.~\cite{Bandyopadhyay:2007kx}.
If future cosmological data finds a positive signal above $0.1$ eV it would be impossible to measure unambiguously the hierarchy with cosmological information only. However, if such a signal is found, direct neutrino mass searches, as tritium beta decay experiments, could infer the mass of the lightest neutrino and provide the key to determine the hierarchy and leptonic CP violation. Moreover, if the neutrino mass hierarchy turns out to be inverted, next generation neutrino-less double beta decay experiments could extract
the neutrino character (for a review of the expected sensitivities, see
\cite{Strumia:2006db}). A Majorana neutrino character, together with a non-vanishing leptonic CP violation would point to leptogenesis as the responsible mechanism for the matter-antimatter asymmetry of the universe.

\begin{table*}
\begin{center}
\begin{tabular}{lllll}
\multicolumn{5}{c}{95\% upper limits on $\sum m_{\nu}$ from the literature}\\
\hline
model & base dataset & U.L. (eV) & sys. errors & Ref.\\ 
\hline
$\Lambda$CDM & WMAP5 & 1.3 &  \cite{Kristiansen/Eriksen/Elgaroy:2006} & \cite{dunkley/etal:2009} \\
$w$CDM & WMAP5 & 1.5 &  \cite{Kristiansen/Eriksen/Elgaroy:2006} &\cite{dunkley/etal:2009} \\
\hline
$\Lambda$CDM & WMAP5$+$BAO$+$SN & 0.67 & - & \cite{komatsu/etal:2009} \\
$w$CDM & WMAP5$+$BAO$+$SN & 0.80 & - & \cite{komatsu/etal:2009} \\
\hline
$\Lambda$CDM & WMAP5+$P_{halo}(k)$ & 0.62 & marginalized & \cite{reid/etal:2009} \\
\hline
$w$CDM & WMAP5$+$BAO$+$SN+& & \\
& {\em Chandra} Clusters
 & 0.33 & $\pm 0.1$ & \cite{vikhlinin/etal:2009} \\
$\Lambda$CDM & WMAP3+WL mass function & 1.43 & - & \cite{kristiansen/elgaroy/dahle:2007}\\
\hline
$\Lambda$CDM & WMAP3+Ly-$\alpha$+CMB small&&\\
&+BAO+SN+SDSS \& 2dF $P(k)$ & 0.17 & IGM \cite{bolton/etal:2008} & \cite{seljak/slosar/mcdonald:2006}\\
$\Lambda$CDM & WMAP5+Ly-$\alpha$+CMB small&&\\
+$\alpha$ &+$H_0$ 2001 \cite{freedman/etal:2001} & 0.28 & IGM \cite{bolton/etal:2008} & -\\
\hline
$\Lambda$CDM & WMAP3+$H_0$+$H(z)$ & 0.50  & - & \cite{stern/etal:2009}\\
\hline
$\Lambda$CDM & WMAP5+SN+BAO+WL & 0.54 & +0.04 & \cite{ichiki/takada/takahashi:2009}, \cite{tereno/etal:2009}\\
\hline
$\Lambda$CDM & WMAP5+$b(L)$+$P_{LRG}(k)$ 2006 & 0.28 & bias modeling & \cite{debernardis/etal:2008} \\
$w$CDM & WMAP5+$b(L)$+$P_{LRG}(k)$ 2006 & 0.59 & bias modeling & \cite{debernardis/etal:2008}
\end{tabular}
\caption{\label{table:mnuother} 95\% upper limit on the sum of the neutrino masses $\sum m_{\nu}$ in eV reported in other studies.  All constraints include WMAP, and have been sorted by horizontal lines by the type of secondary probe: geometrical probe(s) BAO and/or SN, galaxy clustering, cluster mass function, Lyman-$\alpha$ forest, cosmic chronometers, weak lensing, and luminosity-dependent galaxy biasing.  In the second Lyman-$\alpha$ analysis listed, $\alpha \equiv dn_s/d ln k$ is also a free parameter.}
\end{center}
\end{table*}
\section{Constraints on the number of relativistic species $N_{\rm rel}$}
\label{nrelsec}
We now consider constraints on the effective number of neutrino species: in this case we assume that neutrino masses are negligible and in the analysis we keep $\sum m_{\nu}$ fixed to $0$.  While in principle both $N_{\rm rel}$ and $\sum m_{\nu}$ should be varied simultaneously, in practice the two are no longer degenerate, and so the assumption $\sum m_{\nu} = 0$ does not change the constraints much \cite{hannestad/raffelt:2006}.

Equation \ref{zequal} defines the major degeneracy between $N_{\rm rel}$ and $\Omega_m h^2$ in the WMAP5 data.  With the strict assumptions of a $\Lambda$CDM cosmology, the measured angular diameter distance can only be recovered if $H_0$ varies with $N_{\rm rel}$, while the $\Omega_m$ and $\Omega_b h^2$ constraints do not relax when $N_{\rm rel}$ is allowed to vary.  This explains the strong degeneracy between $N_{\rm rel}$ and $H_0$ in the WMAP5 posterior distribution.  There is also a weaker degeneracy between $N_{\rm rel}$ and $n_s$.  While WMAP5 tightly constrains the amplitude of fluctuations on very large scales, $k_0 \sim 0.002$ Mpc$^{-1}$, within the WMAP5 degeneracy and assuming a power law primordial spectrum, the allowed linear $P(k)$ pivots about $k_0$ as $N_{\rm rel}$ varies, so that the amplitude of power near the peak at $k \sim 0.03$ Mpc$^{-1}$ (and beyond) increases with $N_{\rm rel}$.  This tilt causes the degeneracy between $N_{\rm rel}$ and $\sigma_8$, and can be constrained both by measuring the galaxy power spectrum shape and constraints on the cluster mass function.  Finally, since $\Omega_m$ and $\Omega_b h^2$ are both tightly constrained, the variation of $H_0$ with $N_{\rm rel}$ induces a degeneracy between $N_{\rm rel}$ and $\Omega_b/\Omega_m$.  This could be probed through gas fractions in clusters or eventually through the amplitude of the BAO signal in large scale structure data.  These well-defined one-dimensional degeneracies would degrade in a less-restricted model of the expansion history, so that further geometrical constraints would be needed to constrain $N_{\rm rel}$ and other parameters like $\Omega_k$ and $w$ simultaneously.

\subsection{$N_{\rm rel}$ in $\Lambda$CDM using several datasets}
\begin{figure}
  \centering
\resizebox{0.7\columnwidth}{!}{\includegraphics{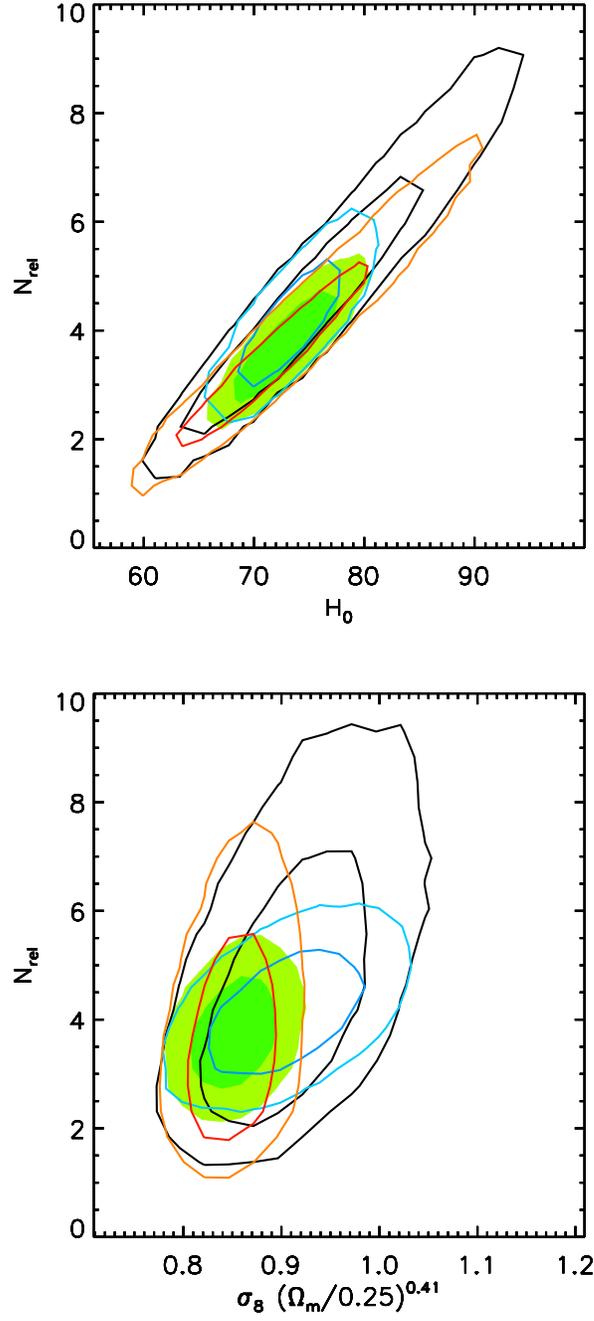}}
  \caption{\label{fig:H0vsNeff} $\Lambda$CDM WMAP5+$P_{halo}(k)$ constraints on $N_{\rm rel}$ vs. $H_0$ and the maxBCG constraint, $\sigma_8 (\Omega_m/0.25)^{0.41}$ (black), and after importance sampling the WMAP5+$P_{halo}(k)$ constraints using the \cite{riess/etal:2009} constraint on $H_0$ (blue), the \cite{rozo/etal:2009} maxBCG cluster constraint on $\sigma_8 (\Omega_m/0.25)^{0.41}$ (red), and both (green).  These contours are similar when the $P_{halo}(k)$ constraints are excluded.}
\end{figure}

Figure \ref{fig:H0vsNeff} shows the degeneracies in the WMAP5+$P_{halo}(k)$ posterior distribution between $N_{\rm rel}$ and both $H_0$ and the maxBCG-constrained parameter $\sigma_8 (\Omega_m/0.25)^{0.41}$; the results are qualitatively the same when the $P_{halo}(k)$ constraints are excluded.  Table \ref{table:Neff2} illustrates constraints on $N_{\rm rel}$ in $\Lambda$CDM for three different base datasets: WMAP5 alone, WMAP5+BAO+SN+\cite{freedman/etal:2001} HST prior ($H_0 = 72 \pm 8$ km s$^{-1}$ Mpc$^{-1}$), and WMAP5+$P_{halo}(k)$ from \cite{reid/etal:2009}.  Both the maxBCG and updated $H_0$ measurement reduce the error on $N_{\rm rel}$ significantly in all three cases.  When both of these constraints are taken together, additional data sets other than WMAP5 do not shift the mean value of $N_{\rm rel}$ or reduce its error: $N_{\rm rel} = 3.76^{+0.63}_{-0.68}$. Therefore the maxBCG and $H_0$ constraints are sufficient to break the key degeneracies in the WMAP5 data in this model.  Since the geometrical constraints from BAO and SN do not contribute significantly to the $N_{\rm rel}$ constraint, they would probably be useful in constraining more general models (i.e., models which vary curvature or the behavior of dark energy) in the context of freely varying $N_{\rm rel}$.

\begin{table*}
\begin{center}
\begin{tabular}{lllll}
\multicolumn{5}{c}{68\% and 95\% confidence intervals for $N_{\rm rel}$}\\
base dataset & - & +maxBCG & $+H_0$ & +maxBCG$+H_0$\\
\hline
&&&&\\
WMAP5 & $5.7^{+2.5}_{-2.4} (^{+4.0}_{-3.9})$ & $4.6^{+1.7}_{-1.6} (^{+3.5}_{-2.6})$ & $4.0^{+1.0}_{-1.0} (^{+2.2}_{-1.8})$ & $3.76^{+0.63}_{-0.68} (^{+1.38}_{-1.21})$ \\
\hline
WMAP5+BAO &&&&\\
+SN+HST & $4.4^{+1.5}_{-1.5} (^{+3.2}_{-2.6})$ & $3.5^{+0.9}_{-0.9} (^{+2.0}_{-1.7})$ & $4.13^{+0.87}_{-0.85} (^{+1.76}_{-1.63})$ & $3.73^{+0.65}_{-0.65} (^{+1.39}_{-1.29})$ \\
\hline
WMAP5&&&&\\
+$P_{halo}(k)$ & $4.8^{+1.8}_{-1.7} (^{+3.9}_{-2.8})$ & $3.9^{+1.3}_{-1.3} (^{+3.1}_{-2.1})$ & $4.16^{+0.76}_{-0.77} (^{+1.60}_{-1.43})$ & $3.77^{+0.67}_{-0.67} (^{+1.37}_{-1.24})$ \\
\hline
\hline
&&&&\\
WMAP5 & $5.7(^{+4.3}_{-4.6})$ & $4.6(^{+2.7}_{-2.3})$ & $4.0(^{+1.6}_{-1.8})$ & $3.76(^{+1.10}_{-1.17})$ \\
\hline
WMAP5+BAO &&&&\\
+SN+HST & $4.4(^{+2.4}_{-2.6})$ & $3.5(^{+1.6}_{-1.8})$ & $4.13(^{+1.57}_{-1.70})$ & $3.73(^{+1.29}_{-1.30})$ \\
\hline
WMAP5&&&&\\
+$P_{halo}(k)$ & $4.8(^{+2.8}_{-3.1})$ & $3.9(^{+2.3}_{-2.2})$ & $4.16(^{+1.40}_{-1.47})$ & $3.77(^{+1.22}_{-1.20})$ \\
\end{tabular}
\caption{\label{table:Neff2} The upper half of the table shows the Bayesian mean, 68\% and 95\% confidence intervals on $N_{\rm rel}$ for a $\Lambda$CDM cosmology.  Here HST refers to the \cite{freedman/etal:2001} constraint on $H_0: 72 \pm 8$ km s$^{-1}$ Mpc$^{-1}$.  When the \cite{riess/etal:2009} $H_0$ constraint is included in columns four and five, the \cite{freedman/etal:2001} HST prior is undone.  As in Table \ref{table:mnu}, in the lower half we compute the maximum $N_{\rm rel}$ for which $\ln L/L_{max} \geq -2$; for ease of comparison we reproduce the central values from the upper portion of the table.  Since the distributions are significantly non-Gaussian unless WMAP, $H_0$, and maxBCG constraints are included, these ``frequentist'' constraints cannot be interpreted as confidence intervals.}
\end{center}
\end{table*}

\subsection{Profile likelihood for $N_{\rm rel}$ in $\Lambda$CDM model}
\label{neffprofilelike}
\begin{figure}
  \centering
 \resizebox{0.7\columnwidth}{!}{\includegraphics{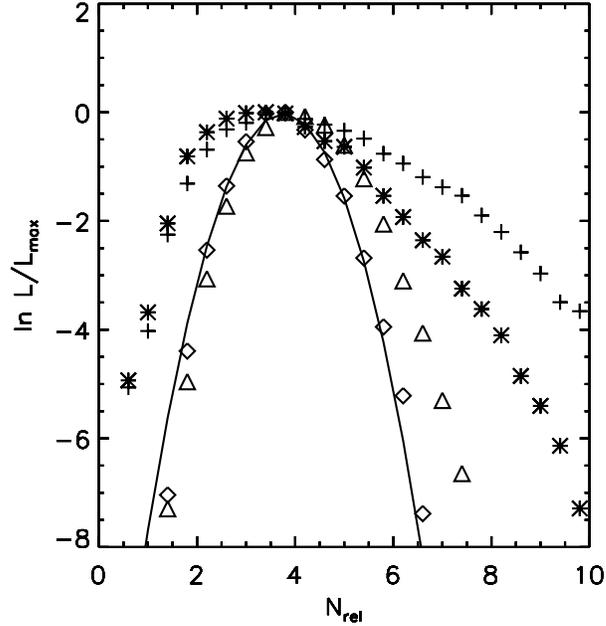}}
  \caption{\label{fig:profilelikenrel}  The profile likelihood defined in Section \ref{method} in bins of $\Delta N_{\rm rel} = 0.4$ for the $\Lambda$CDM model with WMAP5+$P_{halo}(k)$ only (crosses), WMAP5+$P_{halo}(k)$+maxBCG (stars), WMAP5+$P_{halo}(k)$+$H_0$ (triangles), and WMAP5+$P_{halo}(k)$+maxBCG+$H_0$ (diamonds).  The curves are similar when $P_{halo}(k)$ is excluded but $H_0$ and/or maxBCG constraints are included.  The distributions are significantly non-Gaussian, though when both the maxBCG and $H_0$ constraints are included, a Gaussian distribution (solid curve) provides a reasonable fit to the profile likelihood.}
\end{figure}

Figure \ref{fig:H0vsNeff} and Table \ref{table:Neff2} indicate that the likelihood surface is skewed -- contours are more widely separated at large $N_{\rm rel}$.  This finding is in line with the Ref.~\cite{hamann/etal:2007} result that the posterior of $N_{\rm rel}$ only approaches a Gaussian distribution when several datasets are combined.  In Figure \ref{fig:profilelikenrel} this skewness is evident in the profile likelihood for $N_{rel}$.  A Gaussian provides a reasonable fit to the profile likelihood only when both $H_0$ and maxBCG constraints are included.  Therefore, the ``frequentist'' constraints reported in the lower half of Table \ref{table:Neff2} are only meaningful for the last column, which includes $H_0$ and maxBCG constraints.  In that case, there is good agreement with the Bayesian 95\% confidence interval.

\subsection{Comparison with other $N_{\rm rel}$ constraints}
Table \ref{table:Neffother} shows constraints on $N_{\rm rel}$ from other data in the literature.  While WMAP5 provides robust evidence for a lower bound on $N_{\rm rel}$, by itself it cannot provide an upper limit.  The strong correlation between $N_{\rm rel}$ and $H_0$ in the $\Lambda$CDM model makes direct probes of $H_0$ and $H(z)$ the most effective way to constrain $N_{\rm rel}$ with low redshift observations.  The new HST constraint clearly provides a significant improvement over WMAP5 in combination with previously available geometrical constraints (WMAP5+BAO+SN+HST).  The additional constraints on the expansion history from \cite{stern/etal:2009} improve even further, with the WMAP5+$H_0$ error on $N_{\rm rel}$ of 1.0 being reduced to 0.5 with the combination WMAP5+$H_0$+$H(z)$.  The Ly-$\alpha$ constraints are also quite promising as a probe of the power spectrum amplitude at large $k$, but should be reevaluated with the new data and in the context of the possible systematics.  Comparison of measured isotope abundances with BBN calculations provide still  the tightest and completely independent constraints on $N_{\rm rel}$.
\begin{table*}
\begin{center}
\begin{tabular}{llll}
base dataset & $N_{\rm rel}$ & systematic errors & ref.\\
\hline
WMAP5 &  $> 2.3$ (95\% C.L.) & - & \cite{dunkley/etal:2009} \\
WMAP5+BAO+SN+HST & $4.4 \pm 1.5$ & - & \cite{komatsu/etal:2009} \\
\hline
WMAP5+$P_{halo}(k)$ & $4.8^{+1.8}_{-1.7}$ & marginalized & \cite{reid/etal:2009} \\
\hline
$^2$H and $^4$He abundances & $3.18^{+0.22}_{-0.21}$ & reprocessing& \cite{iocco/etal:2009} \\
(BBN) &&contamination&\\
\hline
WMAP3+BAO+SN & $5.3^{+0.4}_{-0.6}$ & IGM \cite{bolton/etal:2008} & \cite{seljak/slosar/mcdonald:2006}\\
+SDSS $P(k)$+2dF $P(k)$ &&&\\
+CMBsmall &&&\\
\hline
WMAP3+$H_0$+$H(z)$ & $4 \pm 0.5$ & - & \cite{stern/etal:2009}
\end{tabular}
\caption{\label{table:Neffother} Constraints on $N_{\rm rel}$ reported in other studies.  Here HST refers to the \cite{freedman/etal:2001} constraint on $H_0: 72 \pm 8$ km s$^{-1}$ Mpc$^{-1}$. All constraints include WMAP except the Big Bang Nucleosynthesis (BBN) constraints, and have been sorted by horizontal lines by the type of secondary probe: geometrical probe(s) BAO and/or SN, galaxy clustering, primordial abundances from BBN, Lyman-$\alpha$ forest, and cosmic chronometers.}
\end{center}
\end{table*}
\section{Conclusions}
\label{conc}
In this paper we have demonstrated that probes of the cluster mass function and increased precision on the Hubble constant can break key degeneracies with CMB observations and yield excellent constraints on both the number of species and sum of the masses of cosmological neutrinos.  When the expansion history is fixed to $\Lambda$CDM, current constraints on $H_0$ \cite{riess/etal:2009} and the cluster mass function \cite{rozo/etal:2009} constrain $\sum m_{\nu} < 0.4$ eV at 95\% confidence.  This bound relaxes to 0.5 eV in the two extended models we considered: wCDM, and the dark coupling model of \cite{gavela/etal:2009}, which allowed curvature, $w$, and coupling strength $\xi$ to vary.  Probing the mass function using X-ray clusters, Ref.~\cite{vikhlinin/etal:2009} combines X-ray cluster data with WMAP5+BAO+SN and finds $\sum m_{\nu} < 0.33$ eV in a $w$CDM cosmology, though the systematic errors still must be clearly quantified.  The optimistic upper bound available from current data, $\sum m_{\nu} \sim 0.3$ eV, almost excludes the scenario in which the neutrino masses are quasi-degenerate.

The constraint on the number of relativistic species $N_{\rm rel}$ from probes of the cluster mass function has not been  widely explored so far.  We find that the combination of WMAP5, \cite{riess/etal:2009} $H_0$, and the maxBCG cluster mass function constraint provides an excellent constraint: $N_{\rm rel} = 3.76^{+0.63}_{-0.68}$.  However, we point out that this constraint does not improve when BAO and SN data are also included; those data sets have been shown to have excellent constraining power on both $\Omega_k$ and $w$ \cite{percival/etal:2009}.  Therefore, we may expect that the constraint on $N_{\rm rel}$ will not relax dramatically if we allow for these additional parameters but include all of these datasets, particularly because of the complementarity of the growth of structure and expansion history constraints we have employed.  These constraints are not competitive with the most optimistic errors from BBN \cite{iocco/etal:2009}, but provide an important consistency check using probes significantly after the BBN epoch.

\section*{Acknowledgments}
B.A.R. acknowledges support from FP7 PEOPLE-2002IRG4-4-IRG and excellent discussions with Steen Hannestad and Risa Wechsler.  LV and RJ acknowledge support from MICINN grant AYA2008-03531. LV acknowledges support of FP7 PEOPLE-2002IRG4-4-IRG\#202182. RJ is supported by a FP7-PEOPLE-IRG grant. OM  is supported by a \emph{Ram\'on y Cajal} contract from MEC, Spain.
\section*{References}
 \bibliographystyle{JHEP}
 \providecommand{\href}[2]{#2}\begingroup\raggedright\endgroup

\begin{thebibliography}{10}

\bibitem{lesgourgues/pastor:2006}
J.~{Lesgourgues} and S.~{Pastor}, {\it {Massive neutrinos and cosmology}},
  {\em \physrep} {\bf 429} (July, 2006) 307--379,
  [\href{http://arXiv.org/abs/arXiv:astro-ph/0603494}{{\tt
  arXiv:astro-ph/0603494}}].

\bibitem{lavacca/bonometto/colombo:2009}
G.~{La Vacca}, S.~A. {Bonometto}, and L.~P.~L. {Colombo}, {\it {Higher neutrino
  mass allowed if Cold Dark Matter and Dark Energy are coupled}},  {\em New
  Astronomy} {\bf 14} (July, 2009) 435--442,
  [\href{http://arXiv.org/abs/0810.0127}{{\tt 0810.0127}}].

\bibitem{hannestad/etal:2007}
S.~{Hannestad}, A.~{Mirizzi}, G.~G. {Raffelt}, and Y.~Y.~Y. {Wong}, {\it
  {Cosmological constraints on neutrino plus axion hot dark matter}},  {\em
  Journal of Cosmology and Astro-Particle Physics} {\bf 8} (Aug., 2007) 15--+,
  [\href{http://arXiv.org/abs/0706.4198}{{\tt 0706.4198}}].

\bibitem{cyburt:2004}
R.~H. {Cyburt}, {\it {Primordial nucleosynthesis for the new cosmology:
  Determining uncertainties and examining concordance}},  {\em \prd} {\bf 70}
  (July, 2004) 023505--+,
  [\href{http://arXiv.org/abs/arXiv:astro-ph/0401091}{{\tt
  arXiv:astro-ph/0401091}}].

\bibitem{cyburt/etal:2005}
R.~H. {Cyburt}, B.~D. {Fields}, K.~A. {Olive}, and E.~{Skillman}, {\it {New BBN
  limits on physics beyond the standard model from \^{}4He}},  {\em
  Astroparticle Physics} {\bf 23} (Apr., 2005) 313--323,
  [\href{http://arXiv.org/abs/arXiv:astro-ph/0408033}{{\tt
  arXiv:astro-ph/0408033}}].

\bibitem{steigman:2007}
G.~{Steigman}, {\it {Primordial Nucleosynthesis in the Precision Cosmology
  Era}},  {\em Annual Review of Nuclear and Particle Science} {\bf 57} (Nov.,
  2007) 463--491, [\href{http://arXiv.org/abs/0712.1100}{{\tt 0712.1100}}].

\bibitem{iocco/etal:2009}
F.~{Iocco}, G.~{Mangano}, G.~{Miele}, O.~{Pisanti}, and P.~D. {Serpico}, {\it
  {Primordial nucleosynthesis: From precision cosmology to fundamental
  physics}},  {\em \physrep} {\bf 472} (Mar., 2009) 1--76,
  [\href{http://arXiv.org/abs/0809.0631}{{\tt 0809.0631}}].

\bibitem{trotta/melchiorri:2005}
R.~{Trotta} and A.~{Melchiorri}, {\it {Indication for Primordial Anisotropies
  in the Neutrino Background from the Wilkinson Microwave Anisotropy Probe and
  the Sloan Digital Sky Survey}},  {\em Physical Review Letters} {\bf 95}
  (June, 2005) 011305--+,
  [\href{http://arXiv.org/abs/arXiv:astro-ph/0412066}{{\tt
  arXiv:astro-ph/0412066}}].

\bibitem{komatsu/etal:2009}
E.~{Komatsu}, J.~{Dunkley}, M.~R. {Nolta}, C.~L. {Bennett}, B.~{Gold},
  G.~{Hinshaw}, N.~{Jarosik}, D.~{Larson}, M.~{Limon}, L.~{Page}, D.~N.
  {Spergel}, M.~{Halpern}, R.~S. {Hill}, A.~{Kogut}, S.~S. {Meyer}, G.~S.
  {Tucker}, J.~L. {Weiland}, E.~{Wollack}, and E.~L. {Wright}, {\it {Five-Year
  Wilkinson Microwave Anisotropy Probe Observations: Cosmological
  Interpretation}},  {\em \apjs} {\bf 180} (Feb., 2009) 330--376,
  [\href{http://arXiv.org/abs/0803.0547}{{\tt 0803.0547}}].

\bibitem{bowen/etal:2002}
R.~{Bowen}, S.~H.~{Hansen}, A.~{Melchiorri}, J.~{Silk}, and R.~{Trotta}, {\it {The impact of an extra background of relativistic particles on the cosmological parameters derived from the cosmic microwave background}}, {\em \mnras} {\bf 334} (Aug., 2002) 760--768, [\href{http://arxiv.org/abs/astro-ph/0110636}{{\tt arXiv:astro-ph/0110636}}].

\bibitem{dunkley/etal:2009}
J.~{Dunkley}, E.~{Komatsu}, M.~R. {Nolta}, D.~N. {Spergel}, D.~{Larson},
  G.~{Hinshaw}, L.~{Page}, C.~L. {Bennett}, B.~{Gold}, N.~{Jarosik}, J.~L.
  {Weiland}, M.~{Halpern}, R.~S. {Hill}, A.~{Kogut}, M.~{Limon}, S.~S. {Meyer},
  G.~S. {Tucker}, E.~{Wollack}, and E.~L. {Wright}, {\it {Five-Year Wilkinson
  Microwave Anisotropy Probe Observations: Likelihoods and Parameters from the
  WMAP Data}},  {\em \apjs} {\bf 180} (Feb., 2009) 306--329,
  [\href{http://arXiv.org/abs/0803.0586}{{\tt 0803.0586}}].

\bibitem{reid/etal:2009}
B.~A. {Reid}, W.~J. {Percival}, D.~J. {Eisenstein}, L.~{Verde}, D.~N.
  {Spergel}, R.~A. {Skibba}, N.~A. {Bahcall}, T.~{Budavari}, M.~{Fukugita},
  J.~R. {Gott}, J.~E. {Gunn}, Z.~{Ivezic}, G.~R. {Knapp}, R.~G. {Kron}, R.~H.
  {Lupton}, T.~A. {McKay}, A.~{Meiksin}, R.~C. {Nichol}, A.~C. {Pope}, D.~J.
  {Schlegel}, D.~P. {Schneider}, M.~A. {Strauss}, C.~{Stoughton}, A.~S.
  {Szalay}, M.~{Tegmark}, D.~H. {Weinberg}, D.~G. {York}, and I.~{Zehavi}, {\it
  {Cosmological Constraints from the Clustering of the Sloan Digital Sky Survey
  DR7 Luminous Red Galaxies}},  {\em ArXiv e-prints} (July, 2009)
  [\href{http://arXiv.org/abs/0907.1659}{{\tt 0907.1659}}].

\bibitem{Verde/etal:2003}
L.~{Verde}, H.~V. {Peiris}, D.~N. {Spergel}, M.~R. {Nolta}, C.~L. {Bennett},
  M.~{Halpern}, G.~{Hinshaw}, N.~{Jarosik}, A.~{Kogut}, M.~{Limon}, S.~S.
  {Meyer}, L.~{Page}, G.~S. {Tucker}, E.~{Wollack}, and E.~L. {Wright}, {\it
  {First-Year Wilkinson Microwave Anisotropy Probe (WMAP) Observations:
  Parameter Estimation Methodology}},  {\em \apjs} {\bf 148} (2003) 195--211.

\bibitem{gavela/etal:2009}
M.~B. {Gavela}, D.~{Hernandez}, L.~{Lopez Honorez}, O.~{Mena}, and
  S.~{Rigolin}, {\it {Dark coupling}},  {\em ArXiv e-prints} (Jan., 2009)
  [\href{http://arXiv.org/abs/0901.1611}{{\tt 0901.1611}}].

\bibitem{riess/etal:2009}
A.~G. {Riess}, L.~{Macri}, S.~{Casertano}, M.~{Sosey}, H.~{Lampeitl}, H.~C.
  {Ferguson}, A.~V. {Filippenko}, S.~W. {Jha}, W.~{Li}, R.~{Chornock}, and
  D.~{Sarkar}, {\it {A Redetermination of the Hubble Constant with the Hubble
  Space Telescope from a Differential Distance Ladder}},  {\em \apj} {\bf 699}
  (July, 2009) 539--563, [\href{http://arXiv.org/abs/0905.0695}{{\tt
  0905.0695}}].

\bibitem{rozo/etal:2009}
E.~{Rozo}, R.~H. {Wechsler}, E.~S. {Rykoff}, J.~T. {Annis}, M.~R. {Becker},
  A.~E. {Evrard}, J.~A. {Frieman}, S.~M. {Hansen}, J.~{Hao}, D.~E. {Johnston},
  B.~P. {Koester}, T.~A. {McKay}, E.~S. {Sheldon}, and D.~H. {Weinberg}, {\it
  {Cosmological Constraints from the SDSS maxBCG Cluster Catalog}},  {\em ArXiv
  e-prints} (Feb., 2009) [\href{http://arXiv.org/abs/0902.3702}{{\tt
  0902.3702}}].

\bibitem{koester/etal:2007}
B.~P. {Koester}, T.~A. {McKay}, J.~{Annis}, R.~H. {Wechsler}, A.~{Evrard},
  L.~{Bleem}, M.~{Becker}, D.~{Johnston}, E.~{Sheldon}, R.~{Nichol},
  C.~{Miller}, R.~{Scranton}, N.~{Bahcall}, J.~{Barentine}, H.~{Brewington},
  J.~{Brinkmann}, M.~{Harvanek}, S.~{Kleinman}, J.~{Krzesinski}, D.~{Long},
  A.~{Nitta}, D.~P. {Schneider}, S.~{Sneddin}, W.~{Voges}, and D.~{York}, {\it
  {A MaxBCG Catalog of 13,823 Galaxy Clusters from the Sloan Digital Sky
  Survey}},  {\em \apj} {\bf 660} (May, 2007) 239--255,
  [\href{http://arXiv.org/abs/arXiv:astro-ph/0701265}{{\tt
  arXiv:astro-ph/0701265}}].

\bibitem{johnston/etal:2007}
D.~E. {Johnston}, E.~S. {Sheldon}, R.~H. {Wechsler}, E.~{Rozo}, B.~P.
  {Koester}, J.~A. {Frieman}, T.~A. {McKay}, A.~E. {Evrard}, M.~R. {Becker},
  and J.~{Annis}, {\it {Cross-correlation Weak Lensing of SDSS galaxy Clusters
  II: Cluster Density Profiles and the Mass--Richness Relation}},  {\em ArXiv
  e-prints} (Sept., 2007) [\href{http://arXiv.org/abs/0709.1159}{{\tt
  0709.1159}}].

\bibitem{brandbyge/etal:2008}
J.~{Brandbyge}, S.~{Hannestad}, T.~{Haugb{\o}lle}, and B.~{Thomsen}, {\it {The
  effect of thermal neutrino motion on the non-linear cosmological matter power
  spectrum}},  {\em Journal of Cosmology and Astro-Particle Physics} {\bf 8}
  (Aug., 2008) 20--+, [\href{http://arXiv.org/abs/0802.3700}{{\tt 0802.3700}}].

\bibitem{spergel/etal:2003}
D.~N. {Spergel}, L.~{Verde}, H.~V. {Peiris}, E.~{Komatsu}, M.~R. {Nolta}, C.~L.
  {Bennett}, M.~{Halpern}, G.~{Hinshaw}, N.~{Jarosik}, A.~{Kogut}, M.~{Limon},
  S.~S. {Meyer}, L.~{Page}, G.~S. {Tucker}, J.~L. {Weiland}, E.~{Wollack}, and
  E.~L. {Wright}, {\it {First-Year Wilkinson Microwave Anisotropy Probe (WMAP)
  Observations: Determination of Cosmological Parameters}},  {\em \apjs} {\bf
  148} (Sept., 2003) 175--194.

\bibitem{spergel/etal:2007}
D.~N. {Spergel}, R.~{Bean}, O.~{Dor{\'e}}, M.~R. {Nolta}, C.~L. {Bennett},
  J.~{Dunkley}, G.~{Hinshaw}, N.~{Jarosik}, E.~{Komatsu}, L.~{Page}, H.~V.
  {Peiris}, L.~{Verde}, M.~{Halpern}, R.~S. {Hill}, A.~{Kogut}, M.~{Limon},
  S.~S. {Meyer}, N.~{Odegard}, G.~S. {Tucker}, J.~L. {Weiland}, E.~{Wollack},
  and E.~L. {Wright}, {\it {Three-Year Wilkinson Microwave Anisotropy Probe
  (WMAP) Observations: Implications for Cosmology}},  {\em \apjs} {\bf 170}
  (June, 2007) 377--408,
  [\href{http://arXiv.org/abs/arXiv:astro-ph/0603449}{{\tt
  arXiv:astro-ph/0603449}}].

\bibitem{lewis/bridle:2002}
A.~{Lewis} and S.~{Bridle}, {\it {Cosmological parameters from CMB and other
  data: A Monte Carlo approach}},  {\em \prd} {\bf 66} (Nov., 2002) 103511.
  
\bibitem{hamann/etal:2007}
J.~{Hamann}, S.~{Hannestad}, G.~G.~{Raffelt}, and Y.~Y.~Y.~{Wong},
{\it {Observational bounds on the cosmic radiation density}}
{\em Journal of Cosmology and Astro-Particle
  Physics} {\bf 8} (Aug., 2007) 21--+,
  [\href{http://arXiv.org/abs/0707.0440}{{\tt
  arXiv:astro-ph/0707.0440}}].

\bibitem{percival/etal:2007}
W.~J. {Percival}, R.~C. {Nichol}, D.~J. {Eisenstein}, J.~A. {Frieman},
  M.~{Fukugita}, J.~{Loveday}, A.~C. {Pope}, D.~P. {Schneider}, A.~S. {Szalay},
  M.~{Tegmark}, M.~S. {Vogeley}, D.~H. {Weinberg}, I.~{Zehavi}, N.~A.
  {Bahcall}, J.~{Brinkmann}, A.~J. {Connolly}, and A.~{Meiksin}, {\it {The
  Shape of the Sloan Digital Sky Survey Data Release 5 Galaxy Power Spectrum}},
   {\em \apj} {\bf 657} (Mar., 2007) 645--663,
  [\href{http://arXiv.org/abs/arXiv:astro-ph/0608636}{{\tt
  arXiv:astro-ph/0608636}}].

\bibitem{kowalski/etal:2008}
M.~{Kowalski}, D.~{Rubin}, G.~{Aldering}, R.~J. {Agostinho}, A.~{Amadon},
  R.~{Amanullah}, C.~{Balland}, K.~{Barbary}, G.~{Blanc}, P.~J. {Challis},
  A.~{Conley}, N.~V. {Connolly}, R.~{Covarrubias}, K.~S. {Dawson}, S.~E.
  {Deustua}, R.~{Ellis}, S.~{Fabbro}, V.~{Fadeyev}, X.~{Fan}, B.~{Farris},
  G.~{Folatelli}, B.~L. {Frye}, G.~{Garavini}, E.~L. {Gates}, L.~{Germany},
  G.~{Goldhaber}, B.~{Goldman}, A.~{Goobar}, D.~E. {Groom}, J.~{Haissinski},
  D.~{Hardin}, I.~{Hook}, S.~{Kent}, A.~G. {Kim}, R.~A. {Knop}, C.~{Lidman},
  E.~V. {Linder}, J.~{Mendez}, J.~{Meyers}, G.~J. {Miller}, M.~{Moniez}, A.~M.
  {Mour{\~a}o}, H.~{Newberg}, S.~{Nobili}, P.~E. {Nugent}, R.~{Pain},
  O.~{Perdereau}, S.~{Perlmutter}, M.~M. {Phillips}, V.~{Prasad}, R.~{Quimby},
  N.~{Regnault}, J.~{Rich}, E.~P. {Rubenstein}, P.~{Ruiz-Lapuente}, F.~D.
  {Santos}, B.~E. {Schaefer}, R.~A. {Schommer}, R.~C. {Smith}, A.~M.
  {Soderberg}, A.~L. {Spadafora}, L.-G. {Strolger}, M.~{Strovink}, N.~B.
  {Suntzeff}, N.~{Suzuki}, R.~C. {Thomas}, N.~A. {Walton}, L.~{Wang}, W.~M.
  {Wood-Vasey}, and J.~L. {Yun}, {\it {Improved Cosmological Constraints from
  New, Old, and Combined Supernova Data Sets}},  {\em \apj} {\bf 686} (Oct.,
  2008) 749--778, [\href{http://arXiv.org/abs/0804.4142}{{\tt 0804.4142}}].

\bibitem{Kristiansen/LaVacca/Colombo/Mainini/Bonometto:2009}
J.~R. {Kristiansen}, G.~{La Vacca}, L.~P.~L. {Colombo}, R.~{Mainini}, and S.~A.
  {Bonometto}, {\it {Coupling between cold dark matter and dark energy from
  neutrino mass experiments}},  {\em ArXiv e-prints} (Feb., 2009)
  [\href{http://arXiv.org/abs/0902.2737}{{\tt 0902.2737}}].

\bibitem{LaVaccaKristiansen/Colombo/Mainini/Bonometto:2009}
G.~{La Vacca}, J.~R. {Kristiansen}, L.~P.~L. {Colombo}, R.~{Mainini}, and S.~A.
  {Bonometto}, {\it {Do WMAP data favor neutrino mass and a coupling between
  Cold Dark Matter and Dark Energy?}},  {\em Journal of Cosmology and
  Astro-Particle Physics} {\bf 4} (Apr., 2009) 7--+,
  [\href{http://arXiv.org/abs/0902.2711}{{\tt 0902.2711}}].

\bibitem{LaVacca/Kristiansen:2009}
G.~{La Vacca} and J.~R. {Kristiansen}, {\it {Dynamical Dark Energy model
  parameters with or without massive neutrinos}},  {\em Journal of Cosmology
  and Astro-Particle Physics} {\bf 7} (July, 2009) 36--+,
  [\href{http://arXiv.org/abs/0906.4501}{{\tt 0906.4501}}].

\bibitem{Kristiansen/Eriksen/Elgaroy:2006}
J.~R. {Kristiansen}, H.~K. {Eriksen}, and {\O}.~{Elgar{\o}y}, {\it {Revised
  WMAP constraints on neutrino masses and other extensions of the minimal
  {$\Lambda$}CDM model}},  {\em \prd} {\bf 74} (Dec., 2006) 123005--+,
  [\href{http://arXiv.org/abs/arXiv:astro-ph/0608017}{{\tt
  arXiv:astro-ph/0608017}}].

\bibitem{hamann/etal:2008}
J.~{Hamann}, S.~{Hannestad}, A.~{Melchiorri}, and Y.~Y.~Y. {Wong}, {\it
  {Non-linear corrections to the cosmological matter power spectrum and
  scale-dependent galaxy bias: implications for parameter estimation}},  {\em
  Journal of Cosmology and Astro-Particle Physics} {\bf 7} (July, 2008) 17--+,
  [\href{http://arXiv.org/abs/arXiv:0804.1789}{{\tt arXiv:0804.1789}}].

\bibitem{vikhlinin/etal:2009}
A.~{Vikhlinin}, A.~V. {Kravtsov}, R.~A. {Burenin}, H.~{Ebeling}, W.~R.
  {Forman}, A.~{Hornstrup}, C.~{Jones}, S.~S. {Murray}, D.~{Nagai},
  H.~{Quintana}, and A.~{Voevodkin}, {\it {Chandra Cluster Cosmology Project
  III: Cosmological Parameter Constraints}},  {\em \apj} {\bf 692} (Feb., 2009)
  1060--1074, [\href{http://arXiv.org/abs/0812.2720}{{\tt 0812.2720}}].

\bibitem{mantz/etal:2009}
A.~{Mantz}, S.~W. {Allen}, D.~{Rapetti}, and H.~{Ebeling}, {\it {The Observed
  Growth of Massive Galaxy Clusters I: Statistical Methods and Cosmological
  Constraints}},  {\em ArXiv e-prints} (Sept., 2009)
  [\href{http://arXiv.org/abs/0909.3098}{{\tt 0909.3098}}].

\bibitem{kristiansen/elgaroy/dahle:2007}
J.~R. {Kristiansen}, {\O}.~{Elgar{\o}y}, and H.~{Dahle}, {\it {Using the
  cluster mass function from weak lensing to constrain neutrino masses}},  {\em
  \prd} {\bf 75} (Apr., 2007) 083510--+,
  [\href{http://arXiv.org/abs/arXiv:astro-ph/0611761}{{\tt
  arXiv:astro-ph/0611761}}].

\bibitem{seljak/slosar/mcdonald:2006}
U.~{Seljak}, A.~{Slosar}, and P.~{McDonald}, {\it {Cosmological parameters from
  combining the Lyman-{$\alpha$} forest with CMB, galaxy clustering and SN
  constraints}},  {\em Journal of Cosmology and Astro-Particle Physics} {\bf
  10} (Oct., 2006) 14--+,
  [\href{http://arXiv.org/abs/arXiv:astro-ph/0604335}{{\tt
  arXiv:astro-ph/0604335}}].

\bibitem{bolton/etal:2008}
J.~S. {Bolton}, M.~{Viel}, T.-S. {Kim}, M.~G. {Haehnelt}, and R.~F. {Carswell},
  {\it {Possible evidence for an inverted temperature-density relation in the
  intergalactic medium from the flux distribution of the Ly{$\alpha$} forest}},
   {\em \mnras} {\bf 386} (May, 2008) 1131--1144,
  [\href{http://arXiv.org/abs/0711.2064}{{\tt 0711.2064}}].

\bibitem{stern/etal:2009}
D.~{Stern}, R.~{Jimenez}, L.~{Verde}, M.~{Kamionkowski}, and S.~A. {Stanford},
  {\it {Cosmic Chronometers: Constraining the Equation of State of Dark Energy.
  I: H(z) Measurements}},  {\em ArXiv e-prints} (July, 2009)
  [\href{http://arXiv.org/abs/0907.3149}{{\tt 0907.3149}}].

\bibitem{simon/etal:2005}
J.~{Simon}, L.~{Verde}, and R.~{Jimenez}, {\it {Constraints on the redshift
  dependence of the dark energy potential}},  {\em \prd} {\bf 71} (June, 2005)
  123001--+, [\href{http://arXiv.org/abs/arXiv:astro-ph/0412269}{{\tt
  arXiv:astro-ph/0412269}}].

\bibitem{ichiki/takada/takahashi:2009}
K.~{Ichiki}, M.~{Takada}, and T.~{Takahashi}, {\it {Constraints on neutrino
  masses from weak lensing}},  {\em \prd} {\bf 79} (Jan., 2009) 023520--+,
  [\href{http://arXiv.org/abs/0810.4921}{{\tt 0810.4921}}].

\bibitem{tereno/etal:2009}
I.~{Tereno}, C.~{Schimd}, J.-P. {Uzan}, M.~{Kilbinger}, F.~H. {Vincent}, and
  L.~{Fu}, {\it {CFHTLS weak-lensing constraints on the neutrino masses}},
  {\em ArXiv e-prints} (Oct., 2008) [\href{http://arXiv.org/abs/0810.0555}{{\tt
  0810.0555}}].

\bibitem{seljak/etal:2005}
U.~{Seljak}, A.~{Makarov}, R.~{Mandelbaum}, C.~M. {Hirata}, N.~{Padmanabhan},
  P.~{McDonald}, M.~R. {Blanton}, M.~{Tegmark}, N.~A. {Bahcall}, and
  J.~{Brinkmann}, {\it {SDSS galaxy bias from halo mass-bias relation and its
  cosmological implications}},  {\em \prd} {\bf 71} (Feb., 2005) 043511--+,
  [\href{http://arXiv.org/abs/arXiv:astro-ph/0406594}{{\tt
  arXiv:astro-ph/0406594}}].

\bibitem{debernardis/etal:2008}
F.~{de Bernardis}, P.~{Serra}, A.~{Cooray}, and A.~{Melchiorri}, {\it {Improved
  limit on the neutrino mass with CMB and redshift-dependent halo bias-mass
  relations from SDSS, DEEP2, and Lyman-break galaxies}},  {\em \prd} {\bf 78}
  (Oct., 2008) 083535--+, [\href{http://arXiv.org/abs/0809.1095}{{\tt
  0809.1095}}].

\bibitem{tegmark/etal:2006}
M.~{Tegmark}, D.~J. {Eisenstein}, M.~A. {Strauss}, D.~H. {Weinberg}, M.~R.
  {Blanton}, J.~A. {Frieman}, M.~{Fukugita}, J.~E. {Gunn}, A.~J.~S. {Hamilton},
  G.~R. {Knapp}, R.~C. {Nichol}, J.~P. {Ostriker}, N.~{Padmanabhan}, W.~J.
  {Percival}, D.~J. {Schlegel}, D.~P. {Schneider}, R.~{Scoccimarro},
  U.~{Seljak}, H.-J. {Seo}, M.~{Swanson}, A.~S. {Szalay}, M.~S. {Vogeley},
  J.~{Yoo}, I.~{Zehavi}, K.~{Abazajian}, S.~F. {Anderson}, J.~{Annis}, N.~A.
  {Bahcall}, B.~{Bassett}, A.~{Berlind}, J.~{Brinkmann}, T.~{Budavari},
  F.~{Castander}, A.~{Connolly}, I.~{Csabai}, M.~{Doi}, D.~P. {Finkbeiner},
  B.~{Gillespie}, K.~{Glazebrook}, G.~S. {Hennessy}, D.~W. {Hogg}, {\v
  Z}.~{Ivezi{\'c}}, B.~{Jain}, D.~{Johnston}, S.~{Kent}, D.~Q. {Lamb}, B.~C.
  {Lee}, H.~{Lin}, J.~{Loveday}, R.~H. {Lupton}, J.~A. {Munn}, K.~{Pan},
  C.~{Park}, J.~{Peoples}, J.~R. {Pier}, A.~{Pope}, M.~{Richmond},
  C.~{Rockosi}, R.~{Scranton}, R.~K. {Sheth}, A.~{Stebbins}, C.~{Stoughton},
  I.~{Szapudi}, D.~L. {Tucker}, D.~E.~V. {Berk}, B.~{Yanny}, and D.~G. {York},
  {\it {Cosmological constraints from the SDSS luminous red galaxies}},  {\em
  \prd} {\bf 74} (Dec., 2006) 123507--+,
  [\href{http://arXiv.org/abs/arXiv:astro-ph/0608632}{{\tt
  arXiv:astro-ph/0608632}}].

\bibitem{Kraus:2004zw}
C.~e. {Kraus}, {\it {Final Results from phase II of the Mainz Neutrino Mass
  Search in Tritium $\beta$ Decay}},  {\em Eur.\ Phys.\ J.\ C} {\bf 40} (2005)
  447, [\href{http://arXiv.org/abs/arXiv:hep-ex/0412056}{{\tt
  arXiv:hep-ex/0412056}}].

\bibitem{Albright:2006cw}
C.~H. {Albright} and M.~C. {Chen}, {\it {Model predictions for neutrino
  oscillation parameters}},  {\em \prd} {\bf 74} (2006) 113006,
  [\href{http://arXiv.org/abs/arXiv:hep-ph/0608137}{{\tt
  arXiv:hep-ph/0608137}}].

\bibitem{Bandyopadhyay:2007kx}
{ISS Physics Working Group}, {\it {Physics at a future Neutrino Factory and
  super-beam facility}},  \href{http://arXiv.org/abs/{arXiv:0710.4947}}{{\tt
  {arXiv:0710.4947}}}.

\bibitem{Strumia:2006db}
A.~{Strumia} and F.~{Vissani}, {\it {Neutrino masses and mixings and.}},
  \href{http://arXiv.org/abs/{arXiv:hep-ph/0606054}}{{\tt
  {arXiv:hep-ph/0606054}}}.

\bibitem{freedman/etal:2001}
W.~L. {Freedman}, B.~F. {Madore}, B.~K. {Gibson}, L.~{Ferrarese}, D.~D.
  {Kelson}, S.~{Sakai}, J.~R. {Mould}, R.~C. {Kennicutt}, H.~C. {Ford}, J.~A.
  {Graham}, J.~P. {Huchra}, S.~M.~G. {Hughes}, G.~D. {Illingworth}, L.~M.
  {Macri}, and P.~B. {Stetson}, {\it {Final Results from the Hubble Space
  Telescope Key Project to Measure the Hubble Constant}},  {\em \apj} {\bf 553}
  (May, 2001) 47--72.

\bibitem{hannestad/raffelt:2006}
S.~{Hannestad} and G.~G. {Raffelt}, {\it {Neutrino masses and cosmic radiation
  density: combined analysis}},  {\em Journal of Cosmology and Astro-Particle
  Physics} {\bf 11} (Nov., 2006) 16--+,
  [\href{http://arXiv.org/abs/arXiv:astro-ph/0607101}{{\tt
  arXiv:astro-ph/0607101}}].

\bibitem{percival/etal:2009}
W.~J. {Percival}, B.~A. {Reid}, D.~J. {Eisenstein}, N.~A. {Bahcall},
  T.~{Budavari}, M.~{Fukugita}, J.~E. {Gunn}, Z.~{Ivezic}, G.~R. {Knapp}, R.~G.
  {Kron}, J.~{Loveday}, R.~H. {Lupton}, T.~A. {McKay}, A.~{Meiksin}, R.~C.
  {Nichol}, A.~C. {Pope}, D.~J. {Schlegel}, D.~P. {Schneider}, D.~N. {Spergel},
  C.~{Stoughton}, M.~A. {Strauss}, A.~S. {Szalay}, M.~{Tegmark}, D.~H.
  {Weinberg}, D.~G. {York}, and I.~{Zehavi}, {\it {Baryon Acoustic Oscillations
  in the Sloan Digital Sky Survey Data Release 7 Galaxy Sample}},  {\em ArXiv
  e-prints} (July, 2009) [\href{http://arXiv.org/abs/0907.1660}{{\tt
  0907.1660}}].

\end{thebibliography}
\end{document}